\documentclass[conference]{IEEEtran}
\IEEEoverridecommandlockouts  

\usepackage{lmodern}
\usepackage{cite}
\usepackage{amsmath,amssymb,amsfonts}
\usepackage{algorithmic}
\usepackage{graphicx}
\usepackage{textcomp}
\usepackage{xcolor}
\usepackage{adjustbox}
\usepackage{multirow}
\usepackage{booktabs}
\usepackage{subfigure}
\usepackage{hyperref}
\usepackage[]{bookmark}
\usepackage{flushend}

\def\BibTeX{{\rm B\kern-.05em{\sc i\kern-.025em b}\kern-.08em
    T\kern-.1667em\lower.7ex\hbox{E}\kern-.125emX}}

\newcommand*{\avgalpha}{\overline{\alpha}}

\title{
On the relevance of acoustic measurements 
for creating realistic virtual acoustic environments
\thanks{\hrule\vspace{3pt} This work is in part supported by the DFG  Project No. 444827755.}
}

\author{\IEEEauthorblockN{Siegfried Gündert}
\IEEEauthorblockA{\textit{Akustik} \\ \textit{and Cluster of Excellence Hearing4all} \\
\textit{Carl von Ossietzky Universität}\\
Oldenburg, Germany \\
siegfried.guendert@uol.de}
\and
\IEEEauthorblockN{Stephan D. Ewert}
\IEEEauthorblockA{\textit{Medizinische Physik}\\ \textit{and Cluster of Excellence Hearing4all} \\
\textit{Carl von Ossietzky Universität}\\
Oldenburg, Germany \\
stephan.ewert@uol.de}
\and
\IEEEauthorblockN{Steven van de Par}
\IEEEauthorblockA{\textit{Akustik} \\ \textit{and Cluster of Excellence Hearing4all} \\
\textit{Carl von Ossietzky Universität}\\
Oldenburg, Germany \\
steven.van.de.par@uol.de}
}

\begin{document}

\begin{table*}
  \normalsize{
This work has been submitted to the I3DA 2023 International Conference (IEEE Xplore Digital Library) for possible publication.
Copyright may be transferred without notice, after which this version may no longer be accessible.}
\end{table*}
\newpage

\maketitle
\begin{abstract}
    Geometrical approaches for room acoustics simulation have the advantage of requiring limited computational resources while still achieving a high perceptual plausibility. 
    A common approach is using the image source model for direct and early reflections in connection with further simplified models such as a feedback delay network for the diffuse reverberant tail. 
    When recreating real spaces as virtual acoustic environments 
    using room acoustics simulation, the perceptual relevance of individual parameters in the simulation is unclear. 
    Here we investigate the importance of underlying acoustical measurements and technical evaluation methods to obtain high-quality room acoustics simulations in agreement with dummy-head recordings of a real space. We focus on the role of source directivity.  
    The effect of including measured, modelled, and omnidirectional source directivity in room acoustics simulations was assessed in comparison to the measured reference. 
    Technical evaluation strategies to verify and improve the accuracy of various elements in the simulation processing chain from source, the room properties, to the receiver are presented. Perceptual results from an ABX listening experiment with random speech tokens are shown and compared with technical measures for a ranking of simulation approaches.
\end{abstract}

\begin{IEEEkeywords}
  Room acoustics simulation, source directivity, speech, perceptual evaluation, acoustic measurements
\end{IEEEkeywords}

\section{Introduction}
Various methods for room acoustics simulation and rendering of real reverberant environments are available that allow to generate the Binaural Room Impulse Response (BRIR) for specific source-receiver positions and orientations 
\cite{wendtComputationallyEfficientPerceptuallyPlausibleAlgorithm2014a}, 
\cite{schroderRAVENRealTimeFramework2011}, 
\cite{seeberInteractiveSimulationFreefield2017}. 
Usually, these methods use a simplified model for the room simulation supported by accurate acoustic measurements of source directivity and the receiver's Head Related Impulse Responses (HRIR). 
Such methods then allow to render arbitrary source receiver combinations, dynamic virtual acoustic scenes 
\cite{wendtComputationallyEfficientPerceptuallyPlausibleAlgorithm2014a}, 
\cite{ediss28543} or auralization in architectural planning, something which would be hard to achieve using BRIR measurements of the real space which may even not yet exist. 
For the simulation of existing reverberant environments one challenge is to find the optimal input parameters required by the room simulation method which encompass measurement data of source- and receiver directivity, room geometry, and boundary properties. 
These parameters are available with intrinsic errors due to measurement noise, or geometrical simplifications. 
These errors will accumulate in the simulation and increase deviations from the real environment. 
Moreover, the accuracy required for the individual parameters is still matter of ongoing research \cite{kirschSpatialResolutionLate2021}, \cite{steffensRoleEarlyLate2021}, \cite{starz2022perceptual}, \cite{blauRealisticBinauralAuralizations2021}, \cite{steffensPerceptualRelevanceSpeaker}, \cite{blauASSESSMENTPERCEPTUALATTRIBUTES2018}, 
and can only be assessed meaningfully using perceptual evaluations. 

This study focuses on a comparison of measured BRIRs in a real reverberant environment, with the corresponding simulation of the real environment, by means of technical analyses using various objective room acoustical parameters like reverberation time and direct-to-reverberant ratio, and by means of perceptual evaluations using an ABX test. For the simulation, the room acoustics simulator RAZR is used 
\cite{wendtComputationallyEfficientPerceptuallyPlausibleAlgorithm2014a}, 
\cite{kirschComputationallyEfficientSimulationLate2023}. 
Technical evaluations are performed directly on the BRIRs. For the perceptual evaluation, the BRIRs will be convolved with anechoic source signals to create realistic stimuli of sources being placed in a reverberant environment. More specifically, different methods will be used for measuring, simulating, or equalizing, source directivity in order to get a better understanding of the technical and perceptual requirements for an accurate room simulation and rendering.
All methods use the receiver's HRIRs and the measurement of reverberation time for the estimation of average absorption coefficients yielding homogeneous boundary conditions.
The first method additionally uses accurate acoustic measurements of source directivity.
The second method uses a model of source directivity, based on an extended spherical head model \cite{ewertComputationallyEfficientParametric2021}. 
The third method uses a frequency-independent omnidirectional source directivity. 
Each method is used with and without a direct sound equalization method to compensate for differences between the HRIRs, directionality database, the real 
Head And Torso Simulator, and the loudspeaker used.   

As an evaluation database BRIR measurements of one room with short reverberation time, for multiple source positions and orientations, are presented along with corresponding measurements of primary room geometry, reverberation time, and boundary conditions. 
The design and results of an adapted ABX listening test using random speech tokens in each presentation are shown and discussed.

\section{Room acoustic measurements of a real room}\label{measurements}
\subsection{Measurement setup} \label{rooms}

One room on the campus of Carl von Ossietzky University of Oldenburg was selected to carry out the room acoustics measurements. 
It is an acoustically optimized room used as a communication acoustics lab with moderate absorption and fairly short reverberation time of approx. 0.6 seconds across a broad frequency range.
The room approximately has a shoebox shape of $5.15\times7.05\times2.85~\text{m}^3$. Some walls are equipped with acoustic elements that can be opened to increase or closed to reduce absorption. 
All acoustic elements were closed so that they acted mainly as reflectors and diffusers, hence added some deviation from a pure shoebox shape. 
The floor of the room was equipped with carpet, the ceiling was fully equipped with cylindrical melamine foam absorbers and the 
window front was structured by angled window elements, 
acting as diffusers as well as preventing flutter echo.

\begin{figure}[htbp]
  \centerline{\includegraphics[scale=0.71]{{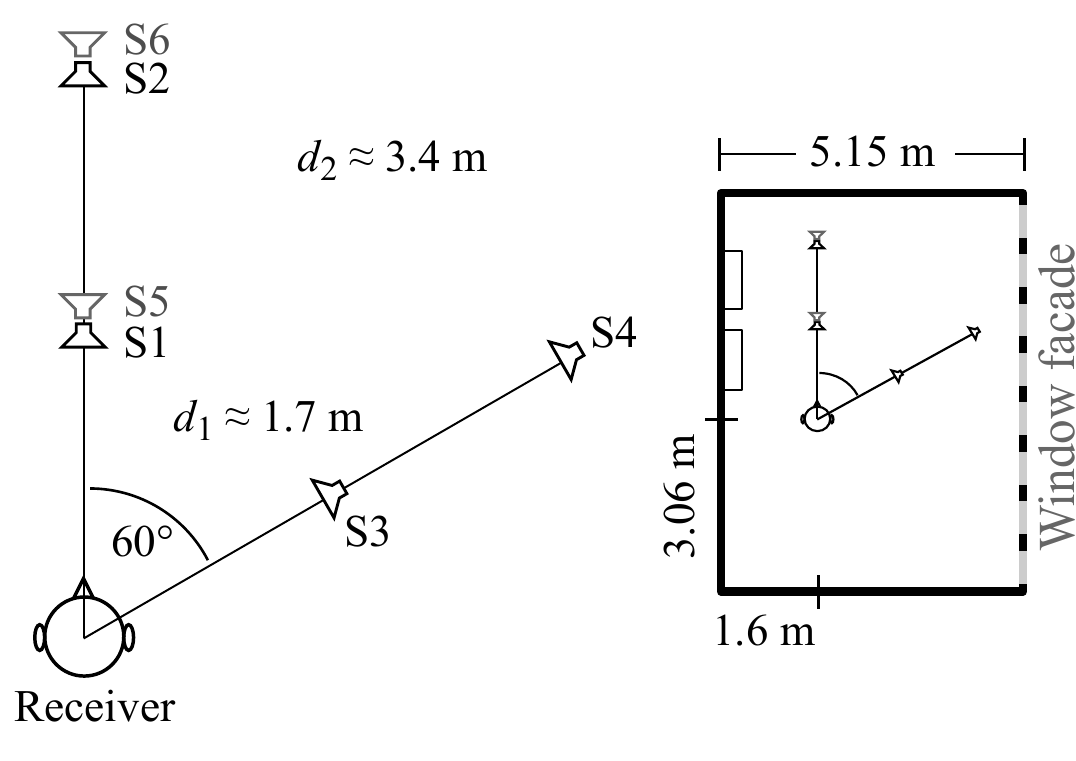}}}
  \caption{Left: Geometrical layout of the sources and receiver. 
  Sources S1 to S4, at a height of 1.3~m, oriented towards the receiver. 
  Source S5 and S6, at a height of 1.5~m, oriented away from the receiver.
  Right: Layout in the room.}
  \label{fig:layout}
\end{figure}

Measurements were performed with different sources and receivers at various positions. 
The geometrical layout of receiver and source positions and orientations was planned to fit in the room as shown in Fig.~\ref{fig:layout}. 
BRIRs were measured using the KEMAR Head And Torso Simulator (HATS) on a fixed receiver position and Genelec 8020 active 2-way monitor loudspeakers at six source positions. 
The loudspeakers were positioned in two different distances (1.7,~3.4)~m and angles (0,~-60)° relative to the receiver position and orientation. 
Four loudspeakers were oriented with the front to the receiver and two loudspeakers in the 0° direction were pointing away from the receiver to investigate the effect of source orientation in room acoustics simulations. 

An omnidirectional source~\cite{kruseOmnidirectionalLoudspeakerBased2013} and a B\&K 4189 omnidirectional 1/2-inch 
microphone were used for the estimation of the reverberation time. 
The available microphone type was free field compensated, and these measurements were only used for deriving technical measures such as the estimation of reverberation time, using backward integration \cite{schroederNewMethodMeasuring} within octave bands.
All impulse response measurements were carried out with an RME~Fireface~UCX audio interface exploiting the highest available precision at a sampling rate of 44.1~kHz.

\subsection{Room Impulse Response measurement method}
All measurements of impulse responses were implemented with the exponential swept sine method as described in 
\cite{novakSynchronizedSweptSineTheory2015} and \cite{novakNonlinearSystemIdentification2010}. 
To get initial estimates of time delays, 
signal-to-noise ratio (SNR), and reverberation time, white noise signals were played back and recorded before each impulse response measurement.
Thereafter, an exponential sweep signal of ten seconds duration from 50~Hz to 22.05~kHz was played back and 
recorded with adapted recording duration, using the estimates based on the preliminary recorded noise signal, to capture all sound until no reverb energy was left. 
The delay estimates were used to create a time window that selected primarily the linear component obtained from the exponential sweep method.

For binaural impulse response measurements, the Genelec 8020 loudspeakers and the KEMAR HATS were used. 
The Genelec 8020 loudspeakers were driven at high sound pressure levels (approx. 86 dB SPL @ 1~m and 1~kHz) without 
corrupting the linear impulse response and without activating the speaker's protection circuitry. 
Both was validated by repeated measurements at different levels and sweep durations. 
Furthermore, Pearson correlation between recordings was checked to be higher than 0.99 for subsequent unprocessed measurements. 
Special care was taken to ensure that binaural information like Interaural Time-arrival Differences (ITD) or Level Differences (ILD) were not altered by the measurement method.
Excerpts of the international speech test signal~\cite{holubeDevelopmentAnalysisInternational2010} were recorded to 
approve that the estimated BRIRs can reproduce the recorded signal by convolution with the playback signal.

\subsection{Boundary properties}\label{alphamethod}
Frequency dependent absorption coefficients of the boundaries were estimated based on Eyring's formula~\cite{eyringREVERBERATIONTIMEDEAD1930}. With this method, one average estimation of absorption coefficients for all boundaries based 
on the room volume $V$, wall surface area $S$, speed of sound $c$ and reverberation time $T_r(f)$ can be calculated according to \eqref{eq:eyringalpha}. 
\begin{equation}
\avgalpha(f) = 1-\exp{\left(\frac{24 \ln{10}}{c} \frac{V}{-S \cdot T_r(f)}\right)}
\label{eq:eyringalpha}
\end{equation}

\section{Simulation of Binaural Room Impulse Responses recreating the real room}

The geometric layout of sources and receivers as described in~\ref{rooms} 
served as reference to create room simulations with the RAZR platform (v0.963, \cite{wendtComputationallyEfficientPerceptuallyPlausibleAlgorithm2014a}, 
\cite{kirschComputationallyEfficientSimulationLate2023}), available as research version at \href{www.razrengine.com}{www.razrengine.com}. 
RAZR uses an image source model (ISM) of shoebox-shaped rooms for the simulation of the early part of the impulse response up 
to a specifiable order of image sources. 
The remaining part of reverb is produced by a feedback delay network (FDN) that collects the ISM output and recursively yields a reverb tail for a selectable number of spatially distributed virtual reverberation sources (VRSs) around the receiver \cite{kirschSpatialResolutionLate2021}. 
Information about geometrical parameters and wall absorption is used in the ISM as well as in the FDN.
Different spatialization and reproduction methods are available whereby this study utilizes the binaural spatialization mode.
Sources and receivers can be placed anywhere in the room as well as oriented by providing spherical coordinates.
Head related impulse responses or a spherical head model can be used for the receiver to create simulated BRIRs. 
Beside an omnidirectional source directivity mode a head shadow model is available as a model to create an approximated source directivity pattern of a human speaker \cite{ewertComputationallyEfficientParametric2021}, \cite{steffensPerceptualRelevanceSpeaker}. 
To allow arbitrary source directivity, a new feature was added to the latest version of RAZR by the authors.
This enables the user to provide source related impulse responses as SOFA file or as Matlab struct. 
A brief suite of tests and validations of direct sound and first reflections in a semi-anechoic scenario and a reverberation room scenario, investigating the expected average spectral power at an omnidirectional receiver using source related impulse responses, was included and is available on the RAZR repository. 

\subsection{Simulation methods}

Three different approaches were developed to recreate the real room with RAZR simulations. 
The approaches mainly addressed the source directivity in RAZR, but another variant of  
modification was the compensation of the direct sound of the simulation to get a direct sound as similar to the measurement as possible if the source and receiver directivity/sensitivity are different to the measurements in the real room.  
It is a reasonable approach to compensate the differences in direct sound since the direct sound is independent of the room acoustics in theory and is available anyway in the binaural impulse response measurement.

All geometrical values as described in \ref{rooms} were used without modification to create the room geometry and the source receiver combinations.
The frequency dependent reverberation time (T30), calculated for octaves from 125~Hz to 8~kHz, was used as input to RAZR. 
These T30 values were then transformed inside RAZR into average absorption coefficients based on the method described in 
\ref{alphamethod} and used for the design of filters inside RAZR.

The receiver was always configured to use the KEMAR database provided by \cite{brarenHighResolutionHeadRelatedTransfer2020}. 
Whereas the database is available at a sample rate of 48~kHz, it was resampled to 44.1~kHz using the SOFAToolbox~(2.1.4).

RAZR was configured to create image sources up to the third order and 24 FDN channels (Virtual Reverberation Sources).
Direct, early, and reverberant parts of the BRIRS were obtained separately as well to be able to filter the direct sound with a compensation filter on the full length signal and synthesize a compensated BRIR afterwards.

Differences between the simulation methods were in source directivity as listed below:

\begin{itemize}
  \item \textbf{Src-Dir}: uses the Genelec 8020 source directivity database from \cite{brinkmannBenchmarkRoomAcoustical2021}.
  \item \textbf{Model-Dir}: uses the head shadow filter model based on \cite{ewertComputationallyEfficientParametric2021} for the source directivity.
  \item \textbf{Omni-Dir}: uses a frequency independent omnidirectional source directivity.
\end{itemize}

When direct sound compensation is applied, the suffix \textbf{DS} is appended.

\subsection{Direct sound compensation method}\label{ch:dscomp}
To reduce differences in direct sound of simulation and measurement, filters and inverse filters were designed from estimates of the measured and simulated direct sound. This was especially important for a source directivity where the direct sound would not fit to the real scenario at all.
The direct sound was extracted by a 3~ms window with von Hann flanks (32 taps) for both, 
the measured and simulated BRIRs with the implication that no reliable information is available below approx. 340~Hz. 
The onset of direct sound can be found by picking the first local minimum in the energy decay curve (EDC, normalized backward integrated 
squared impulse responses), even if the direct sound is 
weaker in amplitude than the early reflections as it is the case for the averted loudspeaker scenarios.
Both direct sound excerpts were used for the design of a compensation FIR filter based on the method in \cite{bolanosAutomaticRegularizationParameter2016}.
When inverting a frequency response function (FRF), small values e.g. notches caused by destructive interference will lead to strong amplifications in the resulting inverse filter. 
Additionally, these small values can have a small SNR, thus noise will increase the error in the affected region in the inverse. 
This is usually an unwanted effect and can be reduced or even removed by including a regularization term  $\beta(f)$ to calculate the inverse filter \ref{eq:bolanos}:

\begin{equation}
  \breve{H}^{-1}(f) = \frac{H^*(f)}{|H(f)|^2+\beta(f)}.
  \label{eq:bolanos}
\end{equation}

\newcommand{\HDS}[1]{H_{\mathrm{DS}\mathrm{#1}}}

Simple approaches define $\beta$ as a constant value thus regularizing frequency regions were no regularization is necessary at all. 
Consequently, the estimated inverse is still containing errors and will not result in something exactly equal to $\breve{H}^{-1}H = I$. 
The chosen approach combined a bandwidth limitation $\alpha(f)$ with notch regularization $\sigma(f)$ where $\beta(f)=\alpha(f)+\sigma(f)^2$. 
Bandwidth was limited by increasing $\alpha(f)$ (here 120 dB) outside a defined pass band region from 20 Hz to 20 kHz where $\alpha=0$.
The $\sigma$ regularization was calculated by the difference of the direct sound magnitude spectrum $\HDS{}$ and a smoothed version using a Savitzky-Golay filter $\HDS{,sm}$ by following case differentiation
\begin{equation}
  \sigma = 
\begin{cases}

  |\HDS{}|-|\HDS{,sm}| & \mathrm{for}~ |\HDS{,sm}|\geq |\HDS{}|  \\
  0 & \mathrm{otherwise.} \\
\end{cases}
\end{equation}
Only specific frequency regions were regularized by this approach. Linear smoothing was chosen because regularization should be narrowband for every frequency region to reduce deviations from the ideal inverse and to prevent broad band modifications of the inverse that could be heard after deconvolution. The compensation step can be expressed by
\begin{equation}
  \HDS{,sim,comp} = \HDS{,sim}\HDS{,meas} \breve{H}_{\mathrm{DS},\mathrm{sim}}^{-1}.\\
\end{equation}
Where DS denotes the direct sound and sim and meas denote simulation and measurement respectively.

\section{Technical evaluation of measured Binaural Room Impulse Responses}
Room acoustic parameters as reverberation time T30, early decay time EDT, direct to reverberant ratio DRR, definition D50, 
and clarity C80 were calculated for all BRIRs at (500, 1k, 2k)~Hz octaves.
Reverberation time T30 was estimated by linear regression from $-5$~dB to $-35$~dB in the energy decay curve (EDC).
Early Decay Time (EDT) was estimated like T30 but in the level range from $-0.1$~dB to $-10$~dB. 
For the estimation of DRR, the windowed direct sound estimates, as described in \ref{ch:dscomp}, were used. 
One important step to achieve comparable DRR values is to use the same window lengths for the reverberant energy calculation of all compared BRIRs.
Interaural time differences were calculated in the time domain from the low pass filtered direct sound impulse responses 
($f_c$=1~kHz, Butterworth 2nd order) by quadratic interpolation between the 3 bins at the maximum of the cross correlation function.
It is an appropriate method for technical evaluation of ITD with subsample precision (subsample delay estimation in frequency domain yielded the same results for these cases, since SNR is high). Whereas perceptually more relevant methods were investigated by \cite{andreopoulouIdentificationPerceptuallyRelevant2017a}.

\section{ABX Listening Test}
The performed ABX listening experiment was a three interval two alternative forced choice matching to sample experiment. 
During this experiment, pairs of corresponding simulated and measured BRIRs ("condition") were compared by convolving each with  dry speech signals. 
The speech signals were taken from the corpus of the Oldenburger Sentence Test (OLSA, \cite{wagener1999entwicklung}) consisting of 120 different sentences of the form name-verb-numeral-adjective-noun. Each interval contained a randomly selected sentence from a single male speaker ensuring that the specific sentences differed between all three intervals. 
This method was chosen to create a more realistic scenario for the perceptual comparison of reverberant environments that focuses on perceptual similarity within one speaker, but not within a specific token of speech.   
Furthermore, all conditions were presented randomly.
Interval A always contained a signal convolved with the simulated BRIR and interval B always contained a signal convolved with the measured BRIR. 
Interval X contained a signal convolved either with the simulated or with the measured BRIR by random choice.
The number of trials per condition was limited to 25. 
The confidence level for the binomial trial model was set to 5\%. When the p-value for one condition was below the chosen 
confidence level, the condition was considered as finished and no further trials were run for that condition.
All convolved signals were normalized to the same loudness using the \mbox{ITU-R BS.1770-4} recommendation to prevent loudness from being a cue.
Sennheiser HD-650 headphones and an RME~Fireface~UC were used as audio playback equipment inside a listening booth.
Calibration was done with a B\&K4153 artificial ear with a microphone capsule B\&K4134, a measuring amplifier B\&K2610, a sound level calibrator 
B\&K4230 and the speech shaped noise corresponding to the OLSA corpus. 
Before every experiment session, a re-check of the calibration was done by a measurement of the
RMS voltage using a true RMS multimeter connected in parallel to the headphones.
All signals were played back at a level of approx. 60~dB SPL (as measured with the artificial ear).
After playback of the three intervals, the following question and task was shown on the graphical user interface (GUI): 
"Which room acoustic environment was in X? Please choose either A or B.".
So the task of the subject was to select A or B (buttons in the GUI), depending on which room acoustic environment 
was perceived by the subject to be the same as X. 
There was no feedback regarding correctness of the subject's answer, and there was no information regarding the momentary condition of the experiment.
The ABX experiment comprised of one block of approximately one hour duration for each of the nine subjects that participated. The duration of the experiment was not fixed and was shorter, 
the better a subject could match the room acoustic environments correctly.
All subjects were normal hearing or with a minor elevation of the hearing threshold that was considered as unimportant for this kind of supra-threshold measurement. 

The data analysis of the ABX experiment is based on the complementary cumulative 
binomial distribution function (CDF) of the given number of trials and number of correct answers for deriving p-values, and based on a signal-detection-theory framework to determine $d'$-values according to \cite{macmillanDetectionTheoryUser2005}, \cite{hautusDecisionStrategiesABX2002} 
and \cite{boleyStatisticalAnalysisABX2009}. 
It was assumed that subjects adopted the differencing decision strategy for deciding their answers, considering that the different conditions were presented in random order. 

\section{Results}

\subsection{Room acoustic measurements}
The reverberation time  estimated from room impulse response measurements with the omnidirectional source and receiver 
is shown in Table~\ref{tab:t30alpha} along with the resulting average absorption coefficients.
The room has a short reverberation time ranging from 0.5~seconds to 0.8~seconds.

\begin{table}[htbp]
  \centering
  \caption{Reverberation time T30 in seconds estimated from measurements with the omnidirectional 
  source and average absorption coefficients derived with \ref{eq:eyringalpha}.}
    \begin{tabular}{|l|r|r|r|r|r|r|r|}
    \cline{2-8}
    \multicolumn{1}{l}{} & \multicolumn{7}{|c|}{Oct. center frequency (Hz)} \\
    \hline
    Param. & 125   & 250   & 500   & 1k  & 2k  & 4k  & 8k \\
    \hline
     T30 (s) & 0,69  & 0,53  & 0,58  & 0,62  & 0,72  & 0,80  & 0,70 \\
    \hline
    $\avgalpha$ (1) & 0,16  & 0,2   & 0,18  & 0,17  & 0,15  & 0,14  & 0,15 \\
    \hline

    \end{tabular}%
  \label{tab:t30alpha}%
\end{table}%

\subsection{Technical evaluation of BRIRs}
\begin{figure}[htbp]
    \centering
    \includegraphics[width=0.52\textwidth]{{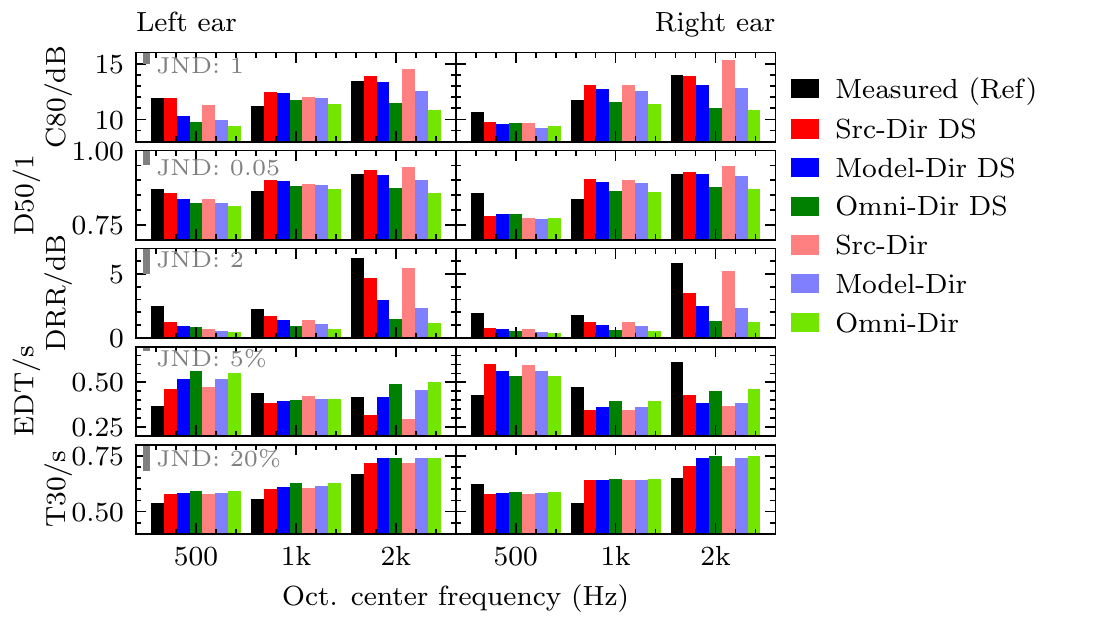}}    
    \caption{Objective room acoustic parameters of BRIRs for source S1 over octave center frequency. The measured reference is shown in black and the colored bars represent the different methods in the same order as the legend. JND values are added to allow for a comparison with perceptual limits.}
    \label{fig:raparam}
\end{figure}

Evaluation results of the objective room acoustic parameters clarity (C80), definition (D50), direct to reverberant ratio (DRR), early decay time (EDT) and T30 are shown in Fig.~\ref{fig:raparam} for the first source position (S1) where the loudspeaker was pointing towards the receiver (KEMAR) at zero degree azimuth. Just noticeable difference (JND) values are added as gray bars with corresponding values in the top left of subplots, to put the data into a perceptual context. JNDs are taken from \cite{DINISO33821}, \cite{larsenMinimumAudibleDifference2008}, and \cite{blevinsQuantifyingJustNoticeable}. Relative JNDs (EDT and T30) were calculated for the average of the visualized value range. One should, however, keep in mind that JNDs can strongly depend on the actual signal that is convolved with a BRIR.

Since reverberation time was one of the parameters for the simulation, it will be described first. The T30 estimates visualized as "Measured (Ref)" are calculated from the binaural measurements and should not be confused with those from omnidirectional measurements being used as input parameters. Since the receivers (left/right ear) and source have a directivity pattern, the resulting reverberation time estimates of the measured BRIRs can deviate from the omnidirectional measurements. Except for the 500~Hz value at the right ear, all T30 "Measured" estimates are below the omnidirectional T30 estimate (see Table~\ref{tab:t30alpha}). The T30 values of simulated BRIRs with method Src-Dir should ideally be near to the values for the measured BRIR as they are simulated using the measured source directivity database. It can be seen that the differences are smaller than the JND for most cases. The values of Omni-Dir are nearest to the omnidirectional measurements as BRIRs are simulated with omnidirectional source directivity. 
T30 estimates of simulated BRIRs are between the T30 estimates of measured BRIRs and omnidirectional RIRs. Source directivity has the strongest influence at 2~kHz. Direct sound compensation did not influence T30 estimates whereas it slightly influenced Early Decay Time (EDT).
Differences for EDT between simulation and measurement are multiples of the JND for most cases. Only left ear Src-Dir at 1~kHz and 2~kHz and Model-Dir DS at 2~kHz diverge less from the measurement.
Direct-to-reverberant ratios (DRR) of measured BRIRs are higher than the DRRs of simulated BRIRs. Direct sound compensation did not reduce differences in DRR, where differences are below JND at 500~Hz and 1~kHz but above at 2~kHz for most conditions except Src-Dir and Src-Dir~DS (left ear).
Definition (D50) ranges below JND for most values but the right ear's 500 Hz values differ up to 0.09 in absolute value. 
Clarity (C80) differences ranges near JND for most values at 500~Hz and 1~kHz, where the differences at 2~kHz are bigger.

The comparisons between measurement and simulation with the objective room acoustic parameters can be used for a ranking of simulation methods. 
This can be done e.g. by calculating relative errors ($1-\textrm{Simulated}/\textrm{Measured}$) between the values of simulated and measured and by averaging the absolute values of relative errors for each method. This results in the ranking (best match first): (1) Src-Dir DS, (2) Src-Dir, (3) Model-Dir DS, (4) Model-Dir, (5) Omni-Dir DS, and (6) Omni-Dir.

Interaural time differences (Table~\ref{tab:itd}) compared to measured did not change with simulation method but with direct sound compensation. Direct sound compensation improved ITD with respect to measured BRIRs. Along the zero degree source positions (S1,2,5,6) ITDs are expected to be 0~µs for 0° azimuth. Estimates of simulated ITDs are near to expectation but the measured ITDs are approx. 30~µs. Differences between ITDs of simulated and measured are smaller for the 60° azimuth positions. 
Just noticeable difference of ITD is highly dependent on stimulus and environment, thus a direct comparison is not expediant. A previous study, \cite{klockgetherJustNoticeableDifferences2016},  shows lowest JNDs of 20~µs in anechoic and approx. 60~µs in echoic conditions for an impulsive snare drum stimulus.

\begin{table}[htbp]
  \centering
  \caption{Interaural time differences ITD (µs) estimated by quadratic interpolation in cross correlation function of low-pass filtered BRIRs. Estimates for measured and simulated with (RAZR DS) and without direct sound compensation (RAZR).}
    \begin{tabular}{|l|r|r|r|r|r|r|}
      \cline{2-7}
      \multicolumn{1}{c}{ITD (µs)} & \multicolumn{6}{|c|}{Source} \\
      \multicolumn{1}{c|}{} & S1 & S2 & S3 & S4 & S5 & S6 \\
      \hline
      Measured & 31    & 27    & 548   & 548   & 34    & 29 \\
      RAZR DS & 31    & 27    & 549   & 550   & 36    & 30 \\
      RAZR & 3     & 3     & 539   & 530   & 5     & 6 \\
    \hline
    \end{tabular}%
  \label{tab:itd}%
\end{table}

\subsection{ABX Listening Test}
\begin{figure}[htbp]
    \centering
    \includegraphics[scale=1]{{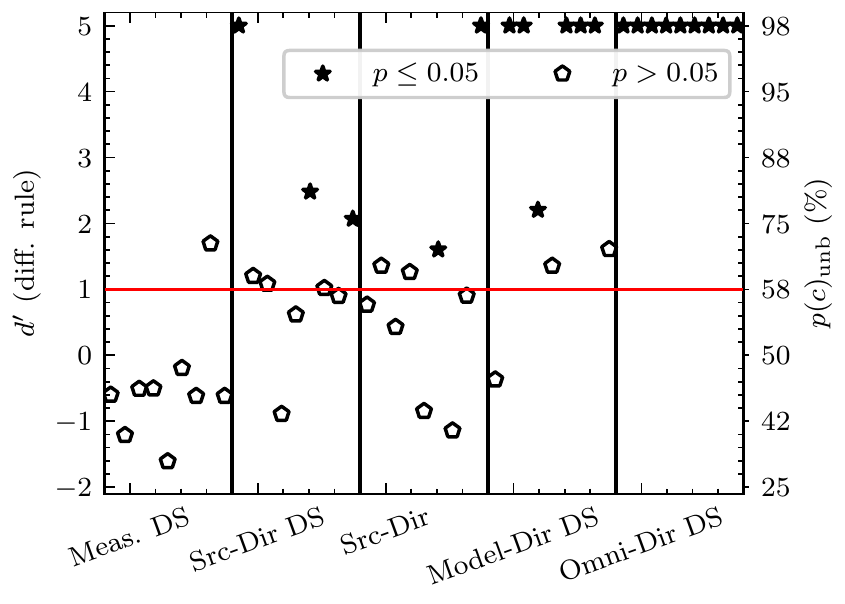}}
    \caption{Results of the ABX listening test. Sensitivity $d'$ is shown on the left  for differencing decision rule sorted over condition and subject. The right ordinate shows unbiased percentage correct values.}
    \label{fig:dprimemk}
\end{figure}

Results of ABX listening tests are shown in Fig.~\ref{fig:dprimemk}. For every condition and subject, $d'$ values based on the differencing decision rule are shown. Filled star markers denote $d'$ values where the experiment outcome, interpreted as binomial trial, has a statistical significance level of 5\% or less (p-value). Pentagons show values where \mbox{p-values} were higher than 5\%. Negative $d'$ with very low percentage of correct ($p(c)$) can also have a complementary probability (binomial CDF) below the chosen 5\%, but no outcomes of this kind were observed. To investigate authenticity of direct sound compensation, another condition ("Meas. DS") was introduced, where the measured BRIR was manipulated by replacing the direct sound with the compensated direct sound from the simulation. The accuracy is somewhat limited towards low frequencies due to the short window that is used to separate direct from reverberant sound components.
For the condition \mbox{Meas. DS}, none of the nine subjects reached a p-value below 5\% within 25 trials, one subject reached a $d'$ value above 1 and eight subjects had a $d'$ below zero. 
Condition \mbox{Src-Dir} was differentiated by two subjects and the remaining seven were below or near to a sensitivity index ($d'$) of 1. 
Three subject differentiated Src-Dir~DS significantly, the remaining six were near or below 1 in sensitivity. The condition \mbox{Model-Dir~DS} was differentiated significantly by six of nine subjects and \mbox{Omni-Dir~DS} was  differentiated by all nine subjects significantly.

\section{Discussion}
The subjective evaluation made in the ABX listening tests indicates that accurate rendering of source directivity is of high importance to achieve close to authentic room simulation. The condition using an omnidirectional source directivity pattern (Omni-Dir DS), which differs strongly from true directivity of the loudspeaker, allowed subjects to consistently hear differences between the measured reference and simulation whereas using measured source directivity data in the simulation (\mbox{Src-Dir DS}, and \mbox{Src-Dir}) made it much more difficult to hear any differences.
The randomization of the selected sentences from the single speaker that were used in each interval of the ABX test prevented subjects from focussing on cues available in a specific stimulus, and made it necessary for subjects to listen to differences relating to source and room characteristics. We will refer to this method as a means to determine "source and room specific authenticity". This differs from the most strict form of (stimulus specific) authenticity that allows to also evaluate differences using a fixed, single stimulus. Using random sentences could lead to two intervals in the ABX test accidentally having very similar properties which could lead a subject to focus specifically on these properties and to use them as cue instead of differences created by differing BRIRs. This scenario could also lead accidentally to negative sensitivity values.
Feedback from the subjects revealed that cues changed with every interval and hence focus changed as well. Mentioned cues were 'localization', 'fricatives or consonants', 'high frequency energy', 'distance' (direct to reverberant energy). However, low frequency energy was not mentioned, which was expected to be one critical cue for comparison of reverberant environments, especially for small to medium-sized rooms (as investigated in this study). Binaural cues in localization may be due to the deviation of the real room from the shoebox approximation leading to differing first reflections (left wall consisted of acoustic elements) that arrive in a time window, where summing location occurs.

For perceptual validation of the direct sound compensation method, the measured BRIR, with replaced direct sound from the direct sound compensated simulation was added to the ABX listening test as a condition (\mbox{Meas. DS}). The outcome of the experiment for this condition indicates that the direct sound compensation was nearly indistinguishable from the real measured direct sound.

In order to obtain a concise evaluation of all conditions measured in the subjective experiment a rank ordering was derived. By arbitrarily choosing $d'>1$ as a criterion for differentiation, the ranking of measured conditions would be as follows (the number of $d'>1$ is shown in parentheses): Meas. DS (1), Src-Dir (4), Src-Dir~DS (6), Model-Dir~DS (8) and Omni-Dir~DS (9). When also requiring conditions to be statistically significant together with $d'>1$ the order remains the same, but the counts change (0, 2, 4, 7, 9).
This implies that source directivity has a strong effect on the authenticity of a reverberant environment. The perceivable effect of simulating with source directivity was approximately a low-pass characteristic acting on the reverberation. Thus, the Omni-Dir case had more energy in high frequencies compared to the measured BRIRs and that was clearly a cue for differentiation. The Model-Dir was already similar to the measured BRIR but still was differentiated by most subjects. An adjustment of the head shadow model towards the actual source characteristic would likely improve the result. It should be noted, that computational complexity is usually much smaller for modelled directivities than for measured FIR databases where for each virtual source database queries and convolution are necessary. Low computational complexity can be required for dynamic virtual acoustic scenes.

Although the technical room acoustic measures show differences to the measurements for most methods these differences are below or in the range of the corresponding JND. Somewhat larger differences were observed for EDT. Since EDT is very sensitive to the early reflections and their energy, one reason could be in the geometrical differences between model and reality. In addition to the resulting deviation in time delays of reflections, also the absorption, amount of scattering, and diffraction (loss of energy in the reflected part) can have an influence on these deviations in EDT.
When the relative errors of room acoustic parameters were used for a ranking of the simulation methods it could be found that this ranking is similar but not the same as the ranking derived from the ABX listening tests. This shows that the use of such technical measures is not a reliable basis for ranking of simulation methods in a perceptual sense. These measures show properties of Impulse Responses that are informed by perceptual relevant aspects, but they do not distinguish between excitation signals for which the JND of a measure may be different as well. Besides this, the weighting across the room acoustic parameters that would be needed to get a ranking similar to a ranking formed from perception based decisions is not known. Furthermore, it is not clear whether other measures would be missing to come up with a result more similar to perception based evaluations. Hence, auralization and subjective evaluations are one important tool for comparison of room acoustic environments and for room acoustics planning. Objective/technical room acoustic parameters offer a tool that helps to make good design decisions or that gives hints where problems or differences occur.

Simulations based on the average absorption coefficients of omnidirectional T30 measurements resulted only in slightly higher reverberation times of BRIRs for all methods including those with measured source related impulse response database (except the 500~Hz band). Measured source directivity did not improve T30 values. For the omnidirectional source simulations (Omni-Dir) T30 estimates of BRIRs are very close to the T30 used as input parameters. This indicates that statistical assumptions in RAZR are in line with expectations for this case. 

Interaural time difference values were different for the methods without direct sound compensation. This indicates that there could be binaural cues helping subjects to differentiate these environments in direct sound depending on the stimulus. These  differences are likely introduced by positioning and orientation errors during measurements. This error could be reduced by more exact positioning and orientation and by checking ITD during the execution of measurement to match the expectations. But the ABX listening tests indicate, that direct sound compensation did not improve for the Src-Dir case.

Finally, we want to remark that we used a relatively compact set of inputs in RAZR to model the room acoustic scenario. Required inputs are the room dimensions, the source and receiver positions and orientations, and a measured room impulse response (reverberation time), which could be substituted with absorption coefficients of room boundaries. In addition to these room related properties, databases were used (which are not room specific) which represent the source and receiver characteristics. In many use cases, we assume it will be easy to determine the room related properties, but more difficult to have access to source and receiver directivity databases. When external databases are used, some differences may occur between the actual equipment that is used and the ones that were used for the directivity databases. To accommodate for these possible differences we found that the direct sound compensation method described in this paper provided a good approach to remediate these differences.

\section{Conclusion}

It can be summarized, that measurements play a crucial role for binaural reproduction of real spaces through room acoustics simulation. The results of the technical and perceptual evaluation point out that source directivity is of great importance. This includes the need of an appropriate reproduction of the real source and receiver e.g. by accurate measurements of source and head related impulse responses. Models of sources (and receivers) would need to be fitted to the real directivities to reduce differences to the reference.
Binaural cues can differ due to the approximated shoebox and homogeneous boundary properties especially because of the deviating early reflections. This could be improved by adding inhomogeneous boundary conditions and more complex geometry to the room acoustic model which in turn can increase computational complexity.
Furthermore, objective technical parameters can be helpful during implementation phase for debugging. Where listening to convolved stimuli is at least as important. This is also true for room acoustics planning, where authentic auralization should play a large role. 
Hence, room acoustic reproductions of real spaces are needed for predictive purposes and for research that makes use of dynamic virtual acoustic scenes in order to relate their results to reality.
The results of this study show that almost indistinguishable simulations of real spaces are achievable with the room acoustics simulator RAZR when considering source and room specific authenticity evaluation.

\bibliographystyle{IEEEtran}
\bibliography{refs}
\end{document}